\documentclass{aa501}
\usepackage{graphics}

\begin{document}

\title{Simultaneous single-pulse observations of radio pulsars}
\subtitle{I. The polarization characteristics of PSR B0329+54}
\author{A. Karastergiou 
	\inst{1}
	\and
	A. von Hoensbroech
	\inst{1}
	\and
	M. Kramer
	\inst{2}
	\and
	D. R. Lorimer
	\inst{3}
	\and
	A. G. Lyne
	\inst{2}
	\and
	O. Doroshenko
	\inst{1}
	\and
	\newline
	A. Jessner
	\inst{1}
	\and
	C. Jordan
	\inst{2}
	\and
	R. Wielebinski
	\inst{1}}

\institute{Max-Planck Institut f\"ur Radioastronomie, Auf dem H\"ugel 69, 53121 Bonn, Germany
	\and 
	Jodrell Bank Observatory, University of Manchester, Macclesfield,
	Chesire SK11 9DL, UK
	\and
	Arecibo Observatory, HC3 Box 53995, Arecibo, Puerto Rico, PR 00612, USA}

\abstract{ 
We present the first results from a programme of multi-frequency simultaneous
single pulse observations carried out as part of the European Pulsar Network.
We detail the main data analysis methods and apply them to simultaneous observations of the
strong pulsar B0329+54 at 1.4 and 2.7 GHz using the Jodrell Bank and Effelsberg
radio telescopes respectively. 
The pulses at different frequencies are highly correlated in their
total intensity, as seen in previous experiments, and generally show
consistent position angles of the linearly polarized component.
In contrast, the 
circularly polarized emission sometimes shows clear differences
between pulses received at different frequencies.  
These results are unexpected and warrant
further follow-up studies to interpret them in the context of the intrinsic
bandwidth of pulsar radiation.
} 

\maketitle

\section{Introduction}\label{obssim}

\begin{figure}[t]
\centerline{
\resizebox{\hsize}{!}{\includegraphics{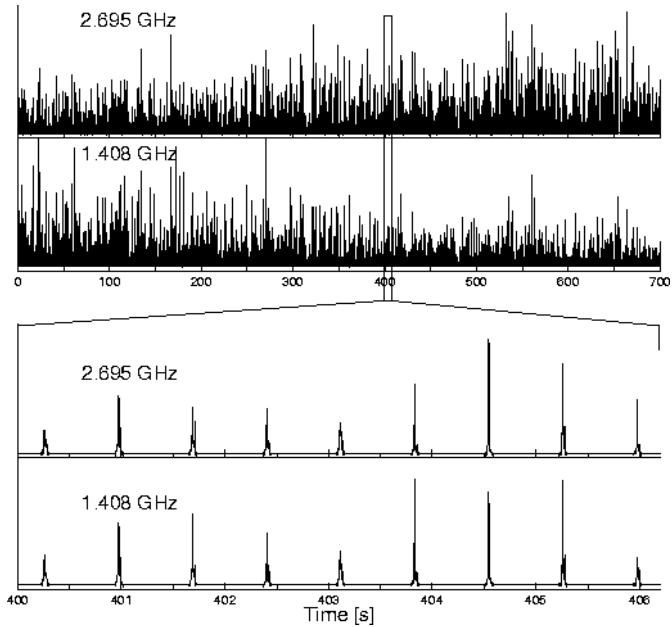}}  
}
\caption[Simultaneous observations of single pulses]
{Total power time series of single pulses of PSR B0329+54 measured at 1.4 GHz
(Jodrell Bank) and 2.7 GHz (Effelsberg). Top two panels show an 11-min
long interval. The long time scale intensity variations between
the two telescopes are caused by interstellar scintillation. Bottom two
panels show a close-up of a few pulses. Off-pulse regions have been
zeroed for clarity. A very good overall correlation is obvious.
}\label{simTP}
\end{figure}

Despite significant observational and theoretical progress since the
discovery of pulsars in 1968, there is still considerable uncertainty
surrounding the detailed physics of the emission process responsible
for the radio pulses observed in the spectrum between 50 MHz all the
way up to 90 GHz (see Melrose 2000 for a review). \nocite{mel00}
Although much of what we understand has been gleaned from studies
of integrated pulse profiles, further understanding of the observed
emission from radio pulsars is perhaps best tackled by studying
individual pulses and their behaviour with time. Any satisfactory
theory of emission has to be able to explain the following observed
properties:

\begin{itemize}
\item moding, drifting and nulling: the relevant phenomena here are 
the pulse-to-pulse patterns that 
consist of
the pulsar profile changing, sub-components of the profile drifting across 
the pulse or the
pulse disappearing altogether for several periods.
\item microstructure: features seen in the pulse on very small time 
scales ($\loa 1$ ms)
\item orthogonal polarization modes: here the polarization position angle 
executes a jump at certain
pulse longitudes, which is often close to $90^{o}$.
\end{itemize}
For further information on these phenomena, see Lyne and Smith (1998)
and references therein.

In the past, a number of experiments have been conducted to
investigate the multi-frequency properties of individual pulses of
radio pulsars (see e.g.~Bartel \& Sieber \cite{bs78}, Boriakoff et
al. \cite{bfs81}, Bartel et al. \cite{bkk81}, Davies et
al. \cite{dls+84}, Kardashev et al. \cite{knn86} and Sallmen et
al. \cite{sb99}). Most of these observations studied total power only
and addressed questions concerning the bandwidth of the emission, the
correlation of microstructure and the validity of the cold plasma
dispersion law for pulsars. Bartel \& Sieber (1978) simultaneously
observed individual pulses of PSR B0329+54 and PSR B1133+16 at 0.3 and 2.7 GHz, using the
Effelsberg radio telescope. The correlation coefficients of the time
series of the pulse energies at the two different frequencies are
found to be around 75\%, indicating
a broadband nature of the emission.
Simultaneous observations between the
Pushchino BSA transit array at 0.1 GHz and the Effelsberg radio
telescope at 1.7 GHz were presented by Bartel et al. (1981) (PSR B0809+74) and
Kardashev et al. (1986) (PSR B0329+54, PSR B0834+06, PSR B1133+16,
PSR B1508+55).
Bartel et al.~(1981)
also showed that nulling occured simultaneously at both frequencies at
0.10 and 1.70 GHz for PSR B0809+74. Simultaneous
observations of the same pulsar by Davies et al.~(1984) at 0.10 GHz,
0.41 GHz and 1.41 GHz using the Lovell telescope at Jodrell Bank 
and the BSA array showed that nulls at 0.4 GHz always corresponded to 
nulls at 0.1 GHz, but not vice versa. Finally, Sallmen et al. (1999)
simultaneously observed giant pulses of the Crab pulsar at 1.4 GHz at
the VLA and 0.6 GHz in Green Bank.
About 70\% of the giant pulses are seen at both 1.4 and 0.6 GHz,
implying a broad emission mechanism bandwidth.

\begin{figure}[t]
\resizebox{\hsize}{!}{\includegraphics{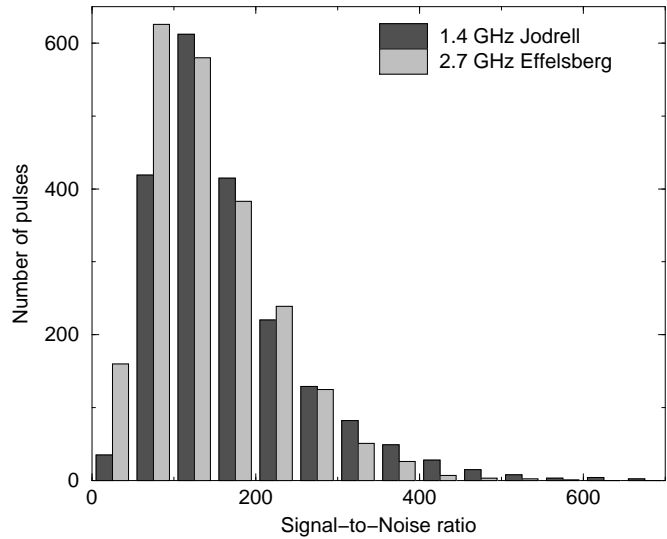}}
\caption[signal-to-noise ratio of simultaneous single pulses of PSR B0329+54]
{
Signal-to-noise ratio distributions of 1912 simultaneously recorded single pulses of PSR B0329+54
between Jodrell Bank (1.4 GHz) and Effelsberg (2.7 GHz). The high quality of the data allows a polarimetric
investigation of the pulses.
}\label{snrhist}
\end{figure}

In 1995, a large scale project was launched by the European Pulsar
Network.  Simultaneous observations of individual pulses were
conducted between various collaborating observatories across Europe
and the world (Effelsberg, Jodrell Bank, Bologna, Westerbork,
Puchshino, Ooty, Torun, Arecibo), as individual experiments.  The data
presented in this paper were obtained during such a polarimetric
experiment between the Jodrell Bank and Effelsberg radio telescopes
observing at the frequencies of 1.41 GHz and 2.70 GHz, respectively.
In this paper, we concentrate on the strong pulsar B0329+54 to
introduce our methods which will be also used in future studies.

In addition to being one of the strongest pulsars in the sky, and
therefore ideal for single-pulse observations, PSR
B0329+54 exhibits many of
the typical pulsar emission features which are not well 
understood. Its profile consists of a core component and at least
four conal outriders surrounding it (e.g.~Kramer 1994). The most
striking feature is perhaps its mode changing property (Lyne 1971, Bartel 
et al.~1982\nocite{bmsh82}),
which is observed at frequencies as high as 32 GHz and 43 GHz
(Kramer et al.~1997)\nocite{kjdw97}.  
The average profile shows significant amounts of linear and
circular emission across the whole pulse. The
position angle (PA) of the average linearly polarized emission shows
frequent orthogonal jumps resulting from orthogonal polarization modes
(OPMs) which are probably simultaneously present (e.g.~McKinnon \&
Stinebring 1998).  Indeed, the seemingly complicated PA swing was
finally unravelled by Gil \& Lyne (1995) who used single pulse
observations at 0.4 GHz to separate the PA swing in two different
curves representing each OPM. 

Despite being studied so well,
simultaneous multi-frequency observations of PSR B0329+54 with full
polarization information were still missing. Since such observations
are essential to distinguish between emission properties caused
by propagation effects (in the pulsar magnetosphere) or those
intrinsic to the emission process (cf.~Melrose 1995), B0329+54 is
an excellent source to begin such a study which is the subject of
the rest of this paper. 
In \S2 we give details about the observations and the telescopes
involved, before we describe the data reduction in \S3.
The results are presented in \S4 and discussed in \S5. We summarize
the main conclusions in \S6 and give an outlook onto further studies in \S7.

\section{Observations}

A sequence of 1912 pulses of very good signal-to-noise ratio (S/N)
were recorded simultaneously between the 76-m Lovell radio telescope
at Jodrell Bank at 1.408 GHz and the 100-m Effelsberg
radio telescope at 2.695 GHz, on the $13^{th}$ of July, 1997.
A dual-channel cryogenic receiver was used on the Lovell telescope,
sensitive to both senses of circular polarization. The system
equivalent flux density was 37 Jy. A $2\times32$-channel filterbank
was used, producing all 4 Stokes parameters for each 1-MHz wide
channel, before
data were de-dispersed in hardware for off-line processing. The effective
time resolution was 158 $\mu$s. 
The Effelsberg data were recorded using a cooled receiver with HEMT
amplifiers providing two circular polarizations with a bandwidth of 80
MHz and a system equivalent flux density of 2.7 Jy.  Given the pulsar's
dispersion measure of 26.776 cm$^{-3}$pc, the smearing time across the
whole bandwidth was 909$\mu$s, compared to a sampling time of 550
$\mu$s. 
Stokes parameters were formed for the full 80 MHz
and transferred to disk. \\
In order to calibrate the data, both
telescopes perform observations of standard unpolarized continuum sources
of known flux density together with the signal of a switched
linearly polarized signal injected into the feeds, as described for
Effelsberg by von Hoensbroech \& Xilouris (\cite{hx97}) and for
Jodrell Bank by Gould \& Lyne (\cite{gl98}).
The average S/N at both observatories was
between 100 and 200, with hardly any pulses below 30 (see
Fig. \ref{snrhist}).

\section{Data analysis}\label{simtech}

Several simultaneous multi-frequency observations of individual pulses
were made on PSR B0329+54 and a number of other pulsars, which will
be described here and in subsequent papers. The data format, starting
time of the observations and time resolution usually differ
between the different telescopes. Therefore the raw data were
calibrated and converted into the common EPN-format (Lorimer et
al. \cite{ljs98}).  Subsequently, a procedure described below was
used, which aligns and re-bins the data to a common reference frame and
temporal resolution.

\subsection{Alignment}

Every pulse emitted from the neutron star does not arrive
simultaneously at different telescopes observing at different
frequencies.  The main reason for this is the interstellar dispersion,
which causes the high frequency radiation of the pulses to arrive
earlier, relative to its lower frequency complement. Further
variations arise from the different path-lengths to the telescopes.
Therefore, in order to align the data from the different telescopes,
it is necessary to transform the site arrival time of the radio
pulses to a common reference frame. This was chosen to be the pulse
arrival time at the solar system barycentre, according to the DE200
ephemeris (Standish \cite{sta82}).  To eliminate dispersion effects,
the time of arrival of each pulse was referred to the time of arrival
of the pulse at infinite frequency as is standard practice in pulsar
timing experiments (see for example Manchester \& Taylor \cite{mt77} 1977).

\begin{figure}[t]
\resizebox{\hsize}{!}{\includegraphics{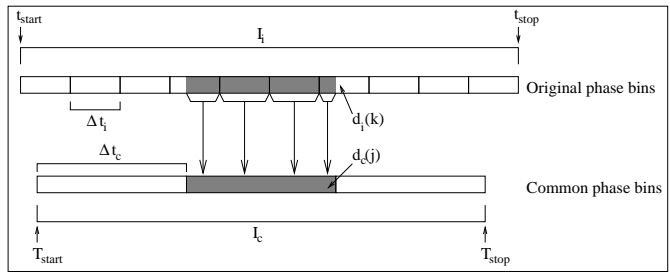}}
\caption[Re-binning procedure]
{Procedure to re-bin data to {\it any} lower time resolution. This 
method is applied for the analysis of simultaneous observations of 
individual pulses. See text for details.
}\label{rebinmethod}
\end{figure}

\subsection{Re-binning}

The calibrated and aligned EPN-files produced from data obtained at
different telescopes  were re-binned to a common time resolution
(Fig.  \ref{rebinmethod}). This is usually the coarsest time
resolution among the original files: A common time interval
$I_c=[T_{\rm start},T_{\rm stop}]$ is defined around each pulse, which
lies within all individual time intervals $I_i=[t_{\rm
start},t_{\rm stop}]_i$ of the different original data files, such
that $I_c \subseteq I_i$, where $i$ denotes each telescope.
The common time interval $I_c$ is divided
into $n_c$ bins of time length $\Delta t_c$. The data contained in all
original bins (time length $\Delta t_i$), which fall into the time
interval of one common bin $\Delta t_c$, are then summed up; those
which fall only partly into this interval are weighted with the
fraction which reaches into $\Delta t_c$. Naming the data of the
$n_c$ common bins $d_c(j),\ j=1..n_c$ and the data contained by the
$n_i$ original bins $d_i(k),\ k=1..n_i$, the re-binned data values can
be computed through
\begin{equation} 
d_c(j)=\frac{\sum_{k=1}^{n_i}d_i(k)\cdot W_j(k)}{\sum_{k=1}^{n_i}W_j(k)}
\end{equation}
with $W_j(k)\in [0:1]$ defining the weight of each original bin $k$ as
the fraction of it which is contained in the common bin $j$.

\begin{figure}[t]
\resizebox{\hsize}{!}{\includegraphics{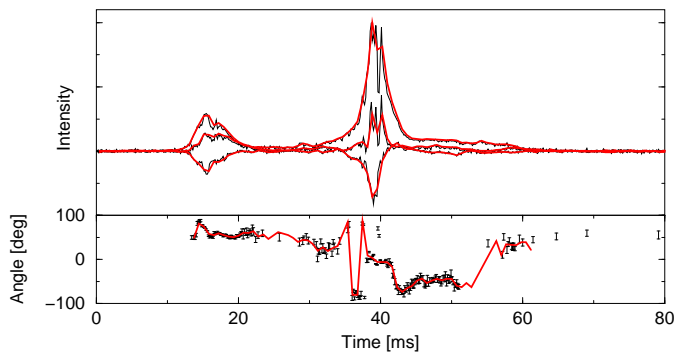}}
\caption[Re-binning example]
{An individual pulse from PSR B0329+54 in full polarization as observed with the 1.4 GHz receiver
in Jodrell Bank (top
panel contains total power, linear and circular polarization, the bottom
one shows the polarization position angle). The thin lines correspond to the original high time
resolution data, the thick lines to the data re-binned to a lower time
resolution using the described routine.
}\label{rebinfig}
\end{figure}

In Fig. \ref{rebinfig} an individual pulse of PSR B0329+54 observed at
1.4 GHz in Jodrell Bank is shown. The thin set of lines corresponds to
the original data, the thick lines represent the re-binned data.
One can see that this procedure affects both the total intensity and the polarization
properties to some extent.
For instance, changing the effective time resolution has an impact on our ability to
detect OPM jumps (notice the bins just before 40 ms). Such effects have been studied by
Gangadhara et al.~(\cite{gxh99}) and will also be addressed elsewhere (Karastergiou et al. \cite{k01}
in prep.).
The circular polarization features particularly studied here are at least
a few bins wide and therefore the rebinning process does not affect our results.
After applying
this re-binning technique, the data entries of the corresponding bins
can be directly compared.

\section{Results}\label{simpairs}

\subsection{Cross-Correlation of Phase Bins}\label{CCbins}

\begin{figure*}
\centerline{
\begin{minipage}{6.5cm} 
\resizebox{\hsize}{!}{\includegraphics{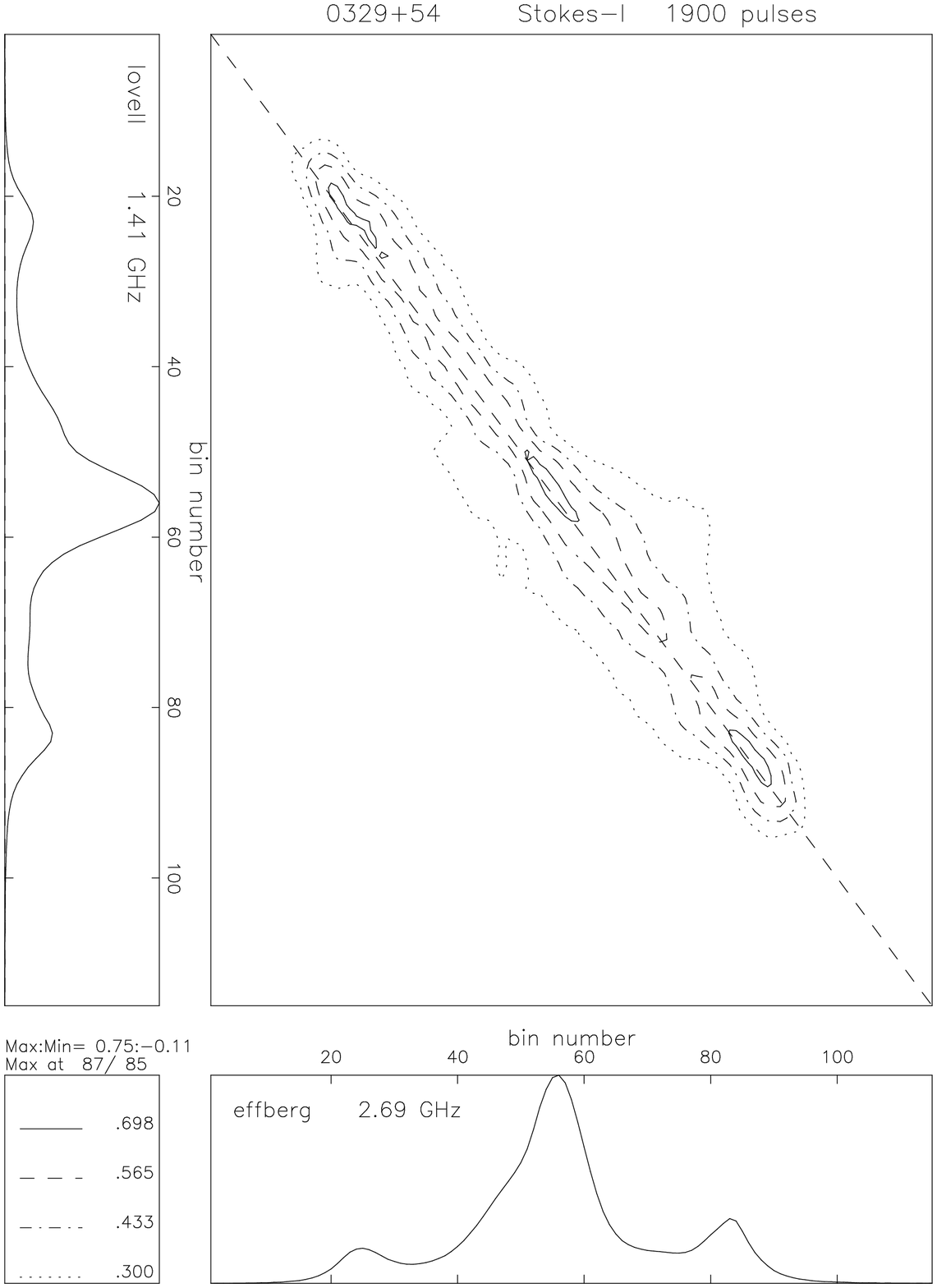}}
\end{minipage}
\begin{minipage}{6.5cm} 
\resizebox{\hsize}{!}{\includegraphics{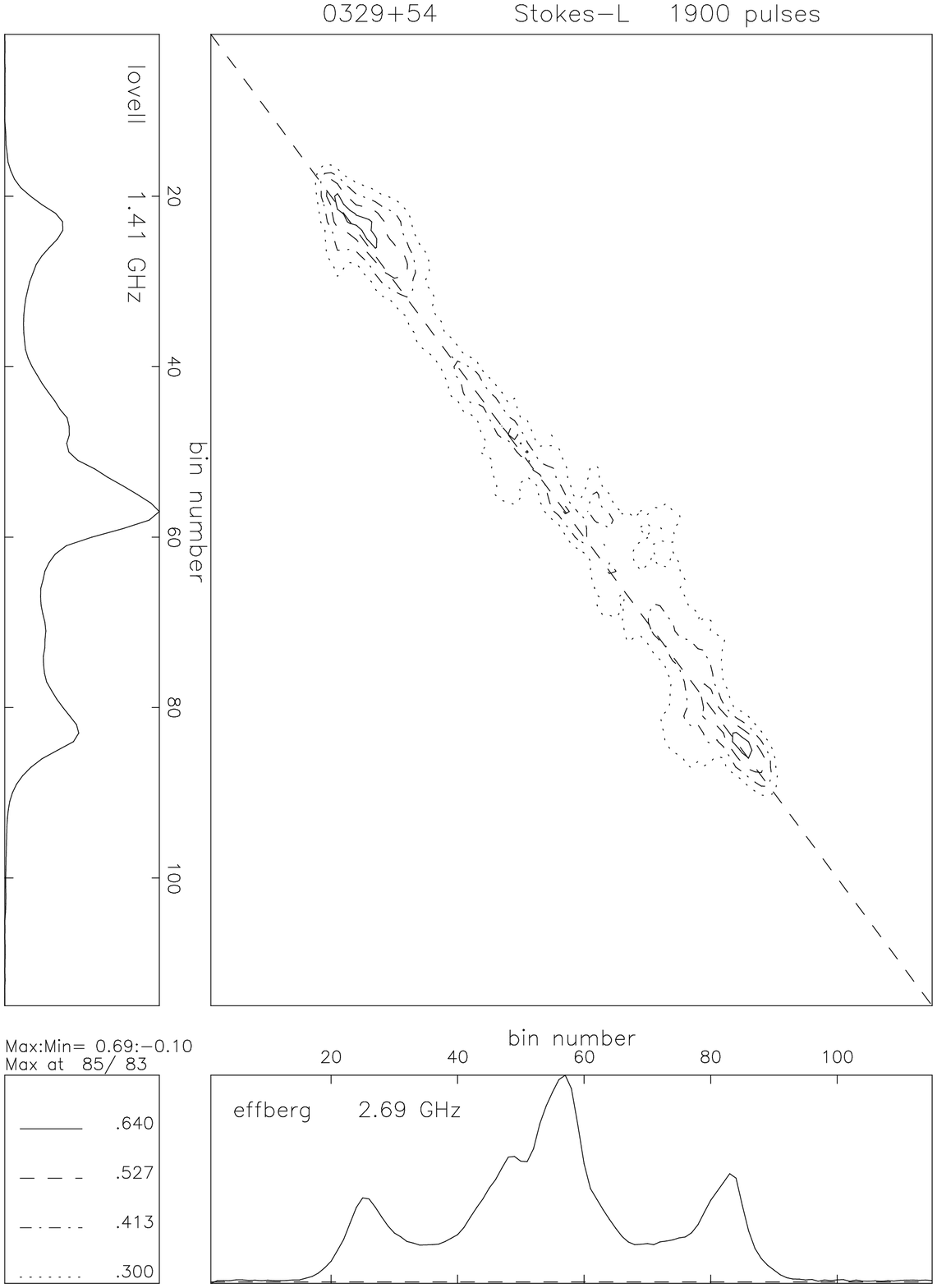}}
\end{minipage}
}
\caption[Bin-by-bin CCFs for I and L]{
Contour plot of a bin-by-bin cross correlation function in total intensity
(left) and linearly polarized power (right). 
The pulse-to-pulse fluctuations of 
each bin at one frequency are correlated against each bin at the other 
frequency. The profiles in the bottom- and left panel represent the
integrated intensities of the respective Stokes parameter. The contour
lines correspond to constant correlation coefficients, their
representations are plotted in the box on the bottom-left.
}\label{CC1}
\end{figure*}

\begin{figure*}
\centerline{
\begin{minipage}{6.5cm} 
\resizebox{\hsize}{!}{\includegraphics{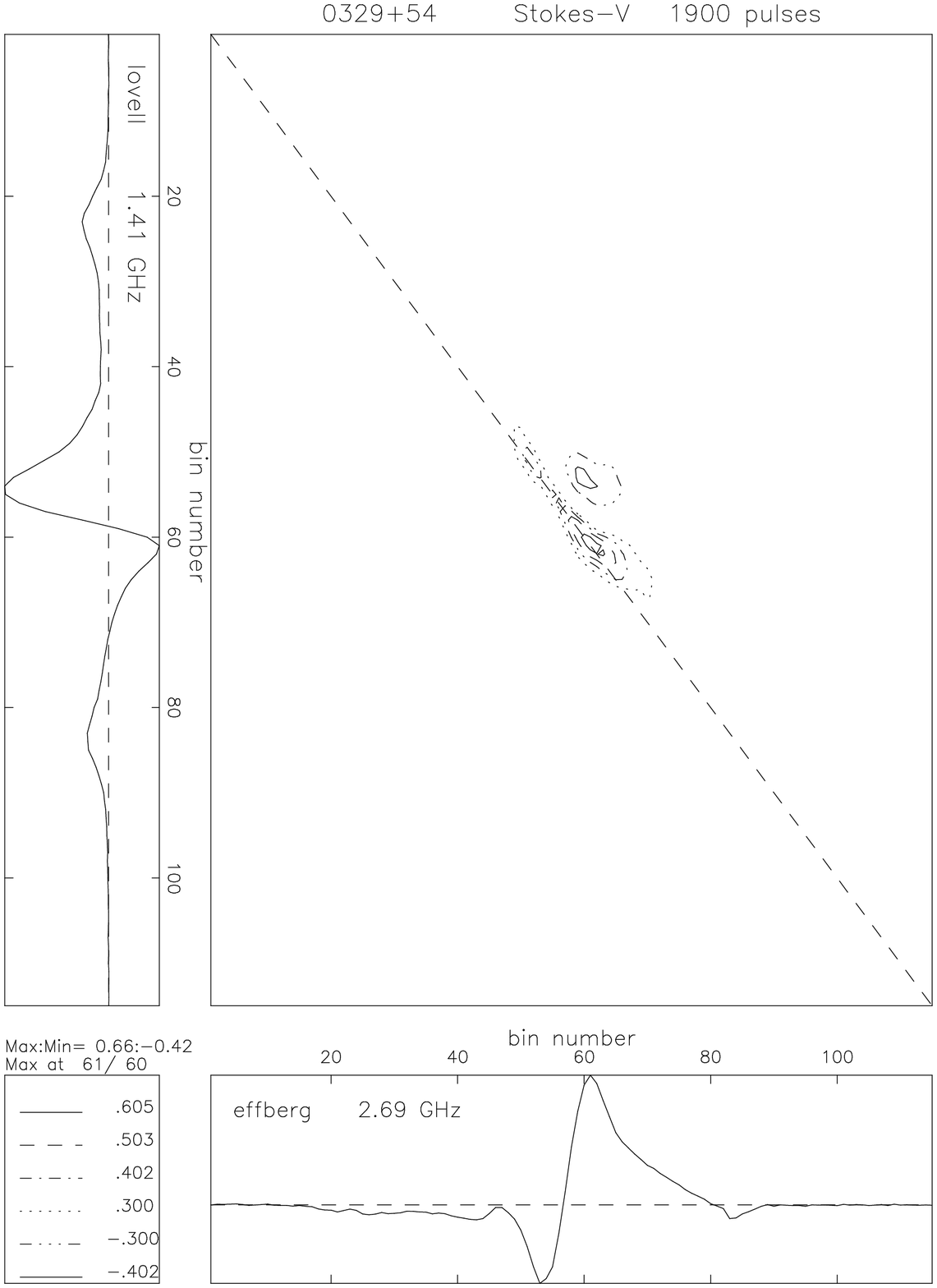}}
\end{minipage}
\begin{minipage}{6.5cm} 
\resizebox{\hsize}{!}{\includegraphics{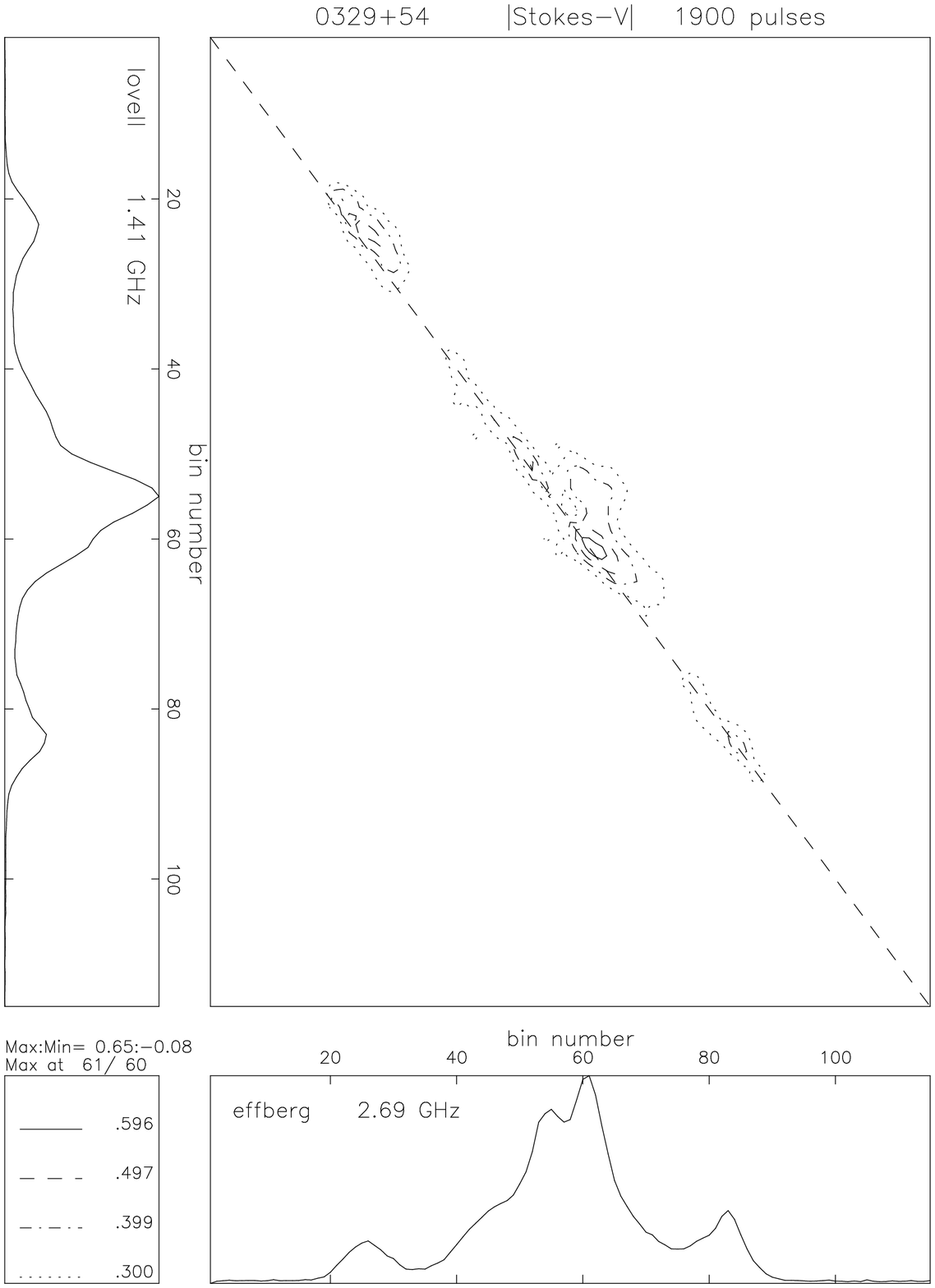}}
\end{minipage}
}
\caption[Bin-by-bin CCFs for V and $|$V$|$]{
Contour plot of a bin-by-bin cross correlation function in circularly- 
(left) and absolute circularly polarized power (right). See caption of
Fig. \ref{CC1} for representations. Note that the little ``island''
above the diagonal in the left plot corresponds to an anti-correlation,
i.e. a disagreement in handedness of the circular polarization. 
The profile for the absolute circular power is calculated by taking the
sum of the absolute circular power of the single pulses and not the
absolute of the sum of the circular power of the single pulses. For this
reason, the two profiles look somewhat different. 
}\label{CC2}
\end{figure*}

In order to study the correlation between the two observing frequencies, 
we conduct a cross-correlation analysis of the pulse-to-pulse
variations of all individual phase bins, following Popov (\cite{p86}).
This yields a two-dimensional
correlation array $c_{i,j}$. Each point
($i$,$j$) is the correlation coefficient between the time series
$f_i(k)$ of bin $i$ at one frequency and the time series $g_j(k)$ of
bin $j$ at the other frequency with $k$ being the pulse number. For
$n$ pulses, $c_{i,j}$ is obtained through
\begin{equation}\label{2dcc} 
c_{i,j}=\frac{1}{n\cdot \sigma_{f,i}\cdot
\sigma_{g,j}}\sum_{k=1}^n[f_i(k)\cdot g_j(k) -<f_i><g_j>]\ ,
\end{equation} 
where $\sigma_{f,i}$ and $\sigma_{g,j}$ represent the {\it rms} deviations of
the time series $f_i(k)$ and $g_j(k)$ respectively.\\ 
Figures \ref{CC1} and \ref{CC2} show contour plots of the correlation
arrays for I, L, V and $|$V$|$. Those points which fall along the
dashed diagonal line represent the correlation array elements $i=j$, i.e.~the
same phase bins between the two frequencies. The width of the
correlated region around the diagonal refers to the time-- and spatial
scale along which the intensities are correlated. The width is largest 
for the total power contour plot.

The distribution of the correlation maxima is quite distinct between the
different polarizations. In total power (Fig. \ref{CC1}, left plot) 
three maxima can be identified, which correspond to the three main
components of the pulse profile. In linear polarization however (Fig.
\ref{CC1}, right plot), the correlation finds its maxima only in the
outer components of the profile. Throughout the middle component---
which is the strongest also in linear polarization, as can be seen from
the integrated profiles on the bottom and the side---the correlation
remains significantly lower.

The circular polarization on the other hand shows the opposite
characteristic. Regarding the ``non-absolute'' plot (Fig. \ref{CC2},
left plot), the maximum is clearly situated in the profile centre. It
is accompanied by a negative ``side maximum'' above the diagonal. This
corresponds to an anti-correlation which is caused by the sense
reversal of the circular polarization often observed at the central
component (also seen in Fig. \ref{indivSP}). 
The other ``side maximum''
expected to be seen beneath the diagonal is not visible due to the
selection of contour lines.  The outer components are completely
uncorrelated. This is mainly caused by heavy frequency dependent
fluctuations in these components and especially the opposite
handedness, which can be frequently observed
(Fig. \ref{indivSP}). This can be obviously seen in the analysis of
the absolute circular polarization (Fig. \ref{CC2}, right plot), where
the difference in handedness is ignored. Some correlation in the outer
profile components is now clearly visible.\\
We note that the results of our cross-correlation
analysis are not significantly affected by the number of pulses included in the study. Dividing
our sample into sufficiently large sub-samples (for this pulsar a stable integrated
profile is obtained after integrating for a few hundred pulses, see
Tab. 1 of Helfand et al.~\cite{hmt75}) does not
change our findings.

\subsection{Individual Pulse Pairs}
\begin{figure*}[t]
\centerline{
\resizebox{12cm}{!}{\includegraphics{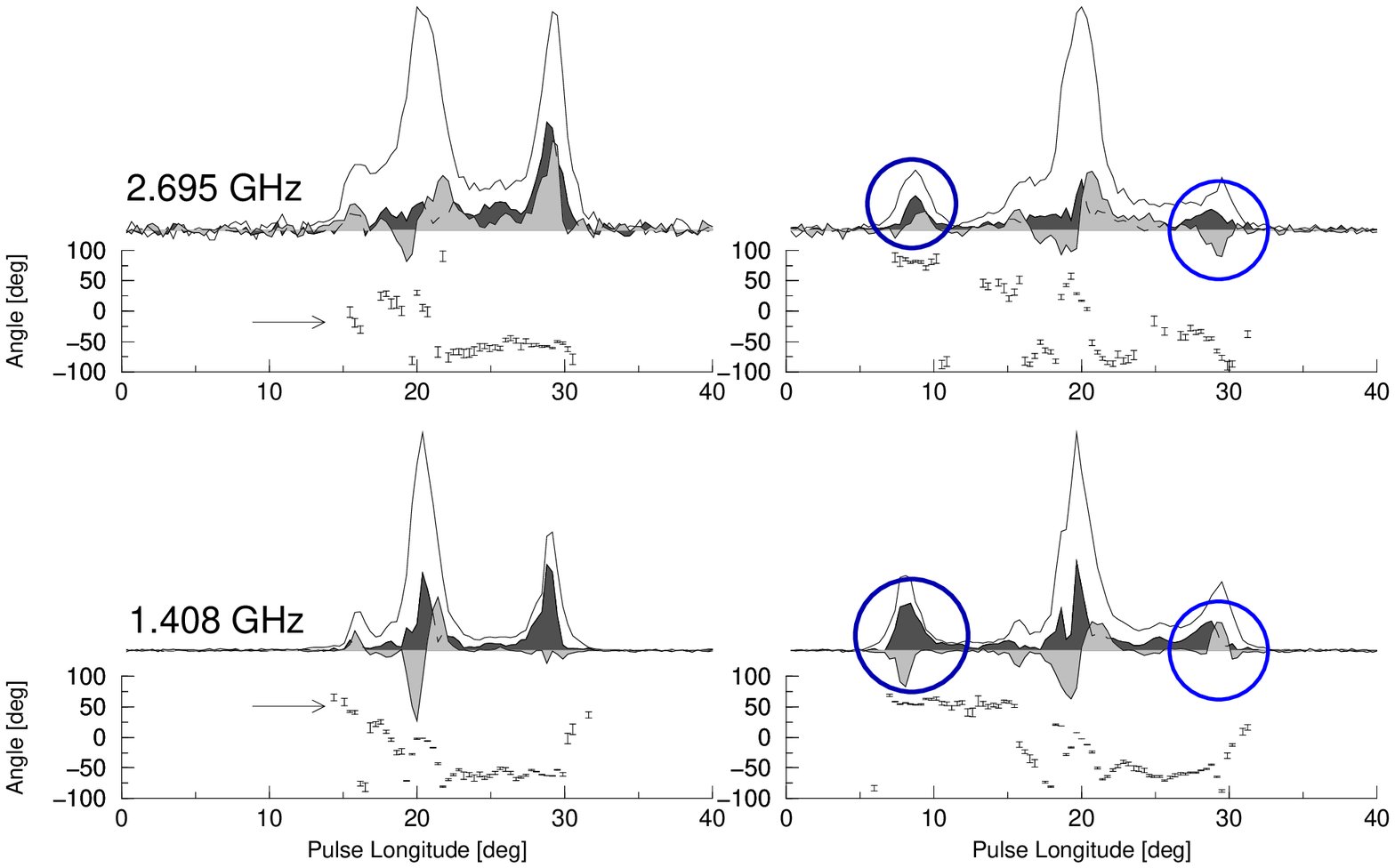}}
}
\caption[Two pulse pairs of PSR B0329+54]
{Two examples of individual pulse pairs of PSR B0329+54, observed in
full polarization. The dark-shaded area represents linearly polarized power and the light-shaded
area circularly polarized power. 
The overall correlation is good, but deviations in the
polarization are larger than those in the
total power. The circular polarization, 
for example, in the wings of the right pulse pair shows a reversed handedness.
Also the arrows in the left pulse pair point to an inconsistency
of the PPA between the two frequencies at a specific pulse longitude.
Note, also,  that only PA points with a significance level of more than 3 $\sigma$ have been plotted.
The pulse components of the higher frequency in the above pulse pairs
appear broader. However, this is not the general case, since we also see
cases where the higher frequency pulse components appear narrower. This fact raises a question
about the spectral indices of pulse components, to be studied in future work.
}
\label{indivSP}
\end{figure*}

To further investigate specific aspects of the cross-correlation analysis
like the disagreement in handedness of the circular polarization, it is often
useful to compare individual pulse pairs. Fig. \ref{indivSP} shows two pulse pairs of
PSR B0329+54. The overall morphology of the pulses is very similar, especially
in total power. However, regarding the polarization characteristics, larger
deviations are observed, as revealed by the cross-correlation plots, especially
concerning the circular polarization but also the polarization position angle.
\begin{itemize}
\item {\it Polarization position angle (PPA)}. Throughout most phases of the pulse longitude, the PPA
is consistent between the two frequencies. Yet at some localized
phases, significant differences between the PPAs are measured (e.g. arrows in
Fig. \ref{indivSP}). These differences indicate the simultaneous presence of both orthogonal
polarization modes at the two frequencies observed. It is worth noting that the PA difference
between the two modes is not always $90^{o}$ as noted also by Gil \& Lyne (\cite{gl95}) and
von Hoensbroech et al. (\cite{hlk}).\\
It should again be noted that determining OPMs is affected by the
temporal resolution of the observation. It is, therefore, quite significant
that we can detect the aforementioned PPA differences after having
rebinned our data (as described in \S3.2).   
\item {\it Circular polarization}. The most prominent differences
observed concern the intensity and handedness of the circularly
polarized component of the radiation. In the trailing component of the
left pulse pair in Fig. \ref{indivSP} 
a high degree of circular polarization is measured
at 2.7 GHz, but none at 1.4 GHz. The cross correlation plot in Fig. \ref{CC2}
indicates that these differences occur preferentially in the outer components,
while the central component is well correlated. 
The fact that the {\em absolute}
circular polarization is well correlated in the outer components indicates
a degree of anti-correlation in the circularly polarized power of the 
outer components.
This is explicitly demonstrated in the right pulse pair 
where the  encircled regions show a handedness that is
opposite between the two frequencies. Note that the handedness is
in fact swapped in both the leading and trailing outer components.
At the same time, however, 
the swing of the circularly polarized power is identical 
at the two frequencies for the central component where we see a
change from negative to positive handedness.
\end{itemize}

\section{Discussion}

We can summarize the results of our analysis as follows:
\begin{itemize}
\item the total intensity is well correlated between the two frequencies
\item linear polarization is correlated  principally in the outer 
components of the profile
\item the two  different orthogonal polarization modes can appear 
simultaneously at the same longitude at different radio frequencies
\item in some parts of the profile the sense of circular polarization 
can change from one frequency to the other while it simultaneously
remains unchanged at other parts of the profile 
\end{itemize}
The good total intensity correlation implies that the mechanism
responsible for emission at one frequency is also responsible for
emission at the other frequency. This strongly suggests that a single
event at the bottom of the pulsar magnetosphere is responsible for a
sub-pulse observed across the spectrum. This can be explained by a
single plasma column streaming along the field lines and radiating
with a large bandwidth. This wide-bandwidth picture seems to be in
some contradiction to a radius-to-frequency mapping (RFM) model which
assumes that the emission at higher frequencies is created closer to
the neutron star than that at lower frequencies. The concept of a RFM
is mainly based on the observations that pulse profiles become narrower
at high frequencies, as expected for pulses emitted in a dipolar
magnetic field (Cordes 1975). It also receives support from the
observations, that the centre of the PA swing lags the profile
midpoint by an amount usually decreasing with frequency, which is
interpreted as a decrease in emission height (Blaskiewicz et
al. \cite{bl91}, von Hoensbroech \& Xilouris \cite{hx97}, Hibschman \& Arons \cite{ha01}). For our
observations, this apparent conflict, for a frequency dependent
emission height and radiation with a bandwidth much larger than the
usual observational spacing, can be solved if the particles creating
the radiation at separate frequencies are part of the same
plasma column.

If the previous assumption is correct, what determines which OPM is
emitted at a given frequency? Are OPMs caused by a propagation effect
like birefringence?  If we accept that pulsar emission is always
radiated in two competing modes of polarization, orthogonal to each
other (McKinnon \& Stinebring \cite{MS98}), at the same time, then the
PA recorded at each pulse phase will have two possible values,
separated by $\approx 90^{o}$. Each set of values forms an S-shaped
curve with pulse phase according to the rotating vector model of
Radhakrishnan \& Cooke (\cite{rc69}), the two curves separated by
$\approx 90^{o}$.  Which of the two PA values is observed would then depend
solely upon which of the competing modes is stronger at every specific
phase. If we could assign a spectral index to the two polarization
modes, which may be different, it would be easy to imagine a situation where one mode
is dominant at the lower frequency and the other at the higher. 

If this scenario of competing OPMs with different spectral indices is correct, the following should be 
observable: At a particular critical frequency, both modes should be present
with almost equal strength and the degree of linear polarization should be
$\approx 0$.  Also, at frequencies at the extreme ends of the observed spectrum
on either side of this critical frequency,
OPMs would only hardly be observed. Although existing observations by
Stinebring et al.~(1984) found some increase in the occurrence of OPMs
at 1.4 GHz compared to 0.4 GHz, this may not be the general case
(Xilouris et al. 1996), and it clearly needs simultaneous observations
of the kind presented in this work to study this interesting question
in detail. In fact, simultaneous polarimetric observations, at three
or more widely spaced frequencies,
could possibly reveal a spectral behaviour of
the OPMs, also to be investigated in future work.

If the two competing OPMs are created by
propagation effects in the pulsar magnetosphere, the polarimetric
differences observed at the two frequencies originate from the
measured radiation having a slightly different history at the two
frequencies. 
The observed changes in the handedness in circular
polarization are hence very intriguing, as they can be indeed most
convincingly explained by some propagation effect.
A change in the
sign of the circular polarization between two frequencies has
been observed in the {\em integrated profiles} of a few pulsars (Han et 
al.~1998), but seeing this for {\em individual pulses} clearly demands
an explanation. The effect however,
seems to be dependent on the pulse longitude, as the central part of
the profile is not affected. 
In the case of PSR B0329+54, the central
component is without doubt a core component (Rankin 1983, Lyne \&
Manchester 1988), which originates from regions near the magnetic
axis. Rankin (1990) suggested that core components are emitted from
altitudes much lower than those for conal components, implying that
the radiation process may be different. Whilst other authors find no
evidence for a difference in the radiation mechanism of core and conal
components (e.g., Lyne \& Manchester 1988, Manchester 1995,
Kramer et al.~1999), one would also
naively expect that emission from a lower altitude should be more
prone to plasma effects as it propagates through the magnetosphere
(see von Hoensbroech et al.~\cite{hlk}). That seems to
be in contrast to what is observed. 

Other factors appear to be more
important than simply the radiation's path length before it escapes the
magnetosphere. For instance, the magnetic field lines in the core
dominated regions have, naturally, a much smaller curvature radius
than the conal regions where the observed changes take place. It is
therefore very tempting to assume that the location of a given profile
longitude mapped on its actual position within the pulsar beam will be
an important parameter in the understanding of OPMs and propagation
effects in the pulsar magnetosphere. Our observations of more pulsars
simultaneously at different frequencies can hence provide very
valuable insight.

\section{Conclusions}

The simultaneous observation of individual pulses in full polarization
allows an investigation of the instantaneous frequency dependence of a
localized emission from the pulsar magnetosphere. The various types of
analyses applied reveal mainly one fact: the polarization
characteristics of single pulses seem to differ a great deal more than
the total power at different frequencies.  This can be seen
in the direct comparison of individual pulses, but also through
various types of cross correlation analyses. It is shown that the
correlation between different frequencies is less 
for the polarized intensities,
especially for the circular polarization. 

It is clearly shown that the differences in the polarization
properties vary between the central and outer components. It seems
that the circular polarization is more stable in the central
component.  In the outer components, the circular polarization
sometimes even shows reversed handedness between the two frequencies.
This implies that polarization fluctuations at a specific longitude
may depend strongly on the curvature of the magnetic field lines at
that longitude mapped on its position within the pulsar beam. This, in
conjunction with the various paths of the radiation through the pulsar
magnetosphere, could be vital in explaining the observed polarization
properties. 

The quality of the new results presented here
underline the necessity of more such observations in the future to
confirm the first results, which should establish multi-frequency
simultaneous observations as a first-class tool for investigating the
emission mechanism of radio pulsars.

\section{Future work}

The present paper is the first in a series of papers concerning simultaneously
observed pulsar data, which focuses on a particular polarimetric experiment between
the Effelsberg and Jodrell Bank radio telescopes. Other observing runs, that have been
carried out within the frame set of the EPN and coordinated by the pulsar group of the
MPIfR in Bonn, will be the subject of papers to 
follow. These runs mainly consist of total-power, simultaneous observations between the
radio-telescopes in Bologna, Westerbork, Torun, Pushchino, Ooty, Effelsberg and Jodrell Bank,
and span through a frequency range between 100 MHz (Pushchino) and 5 GHz (Effelsberg).
We are expecting more polarimetry associated work in the future with the addition of 
Westerbork and the GMRT. Further results from data already taken will
be published in due course.

\acknowledgements

Simultaneous observations of the radio pulsars studied required a maximum degree of organisation and
cooperation between the parties involved. We would especially like to 
thank Christoph Lange and Norbert Wex for their assistance in the realization
of the project. We would also like to acknowledge the scientists who have participated in the
ever growing EPN network, which has at times incorporated the likes of the Pushchino, Ooty and GMRT
radio telescopes. We also want to give credit to all the technical staff at the mentioned observatories
for making our work as easy and as efficient as possible.


\begin{thebibliography}{}

   \bibitem[1986]{ba86} Barnard J. J., Arons J., 1986, ApJ 302, 138

   \bibitem[1978]{bs78} Bartel N., Sieber W., 1978, A\&A 70, 260

   \bibitem[1981]{bkk81} Bartel N., Kardashev N. S., Kuzmin A. D., et al., 1981, A\&A 93, 85

   \bibitem[1982]{bmsh82}
    Bartel N., Morris D., Sieber W., Hankins T.\ H., 1982, ApJ 258, 776

   \bibitem[1991]{bl91} Blaskiewicz M., Cordes J. M., Wasserman I., 1991, ApJ 370, 643 

   \bibitem[1981]{bfs81} Borriakoff V., Ferguson D. C., Slater G., 1981, IAU Symp. No. 95, 199

   \bibitem[1984]{dls+84} Davies J.G., Lyne A.G., Smith F.G., et al., 
   1984, MNRAS 211, 57

   \bibitem[1999]{gxh99} Gangadhara R. T., Xilouris K. M., von Hoensbroech, A., et al., 1999, A\&A 342, 474

   \bibitem[1995]{gl95} Gil J. A., Lyne A. G., 1995, MNRAS 276, L55

   \bibitem[1998]{gl98} Gould D. M., Lyne A. G., 1998, MNRAS 301, 235

   \bibitem[1998]{hmxq98} Han J. L., Manchester R. N., Xu R. X., Qiao G. J., 
   1998, MNRAS 300, 373

   \bibitem[1975]{hmt75} Helfand D. J., Manchester R. N., Taylor J. H.,
   1975, ApJ 198, 661
   \bibitem[2001]{ha01} Hibschman J. A., Arons J., 2001, ApJ 546, 382

   \bibitem[2001]{k01} Karastergiou A. et al., 2001, A\&A, in prep.

   \bibitem[1986]{knn86} Kardeshev N. S., Nikolaev Ya. N., Novikov A. Yu., et al., 1986, A\&A 163, 114

   \bibitem[1994]{kra94} Kramer M., 1994, A\&AS 107, 527

   \bibitem[1997]{kjdw97} Kramer M., Jessner A., Doroshenko O.,
  Wielebinski, R., 1997, ApJ 488, 364

   \bibitem[1999]{kl97} Kramer M., Lange Ch., Lorimer D.R., et al.,
    1999, ApJ 526, 957

   \bibitem[1998]{ljs98} Lorimer D. R., Jessner A., Seiradakis J. H., et al., 1998, A\&AS 128, 541

   \bibitem[1971]{lyn71} Lyne A.G., 1971, MNRAS 153, 27

   \bibitem[1998]{lm88} Lyne A.G., Manchester R.N., 1988, MNRAS 234, 477

   \bibitem[1995]{man95} Manchester, R.N., 1995, JA\&A, 16, 107

   \bibitem[1977]{mt77} Manchester R. N., Taylor J. H., 1977,
  Pulsars, Freeman, San Francisco

   \bibitem[1998]{MS98} McKinnon M. M., Stinebring D. R., 1998, ApJ 502, 883

   \bibitem[1995]{mel95} Melrose D. B., 1995, JA\&A, 16, 137

   \bibitem[2000]{mel00} Melrose D. B., 2000, Pulsar Astronomy - 2000 and Beyond, ASP Conference Series, Vol. 202, 721 


   \bibitem[1986]{p86} Popov M. V., 1986, SOVIET ASTR. 30, 577

   \bibitem[1969]{rc69} Radhakrishnan V., Cooke D. J., 1969, Ap. Letters, 3, 225(RC)

   \bibitem[1983]{ran83} Rankin J. M., 1983, ApJ 274, 358

   \bibitem[1990]{ran90} Rankin J. M., 1990, ApJ 352, 247

   \bibitem[1999]{sb99} Sallmen S., Backer D., Hankins T., Moffett D., Lundgren S., et al., 1999,
      ApJ 517, 460

   \bibitem[1984]{scr84} Stinebring D. R., Cordes J. M., Rankin J. M.,
    Weisberg J. M., Boriakoff, V., 1984, ApJS 55, 247

   \bibitem[1982]{sta82} Standish, E. M., 1982, A\&A 114, 297

   \bibitem[1999]{avh99} von Hoensbroech A., 1999, PhD Thesis, University of Bonn, Bonn

   \bibitem[1998]{hlk} von Hoensbroech A., Lesch H., Kunzl T., 1998, A\&A 336, 209

   \bibitem[1997]{hx97} von Hoensbroech A., Xilouris K. M., 1997, A\&AS 126, 121 
	
   \bibitem[1996]{xkj+96}  Xilouris K. M., Kramer M., Jessner A.,
    Wielebinski R., Timofeev M., 1996, A\&A 309, 481
\end{thebibliography}
\end{document}